\documentstyle[12pt,aasms4]{article}
\begin{document}

\title{Bimodal Accretion Disks: SSD-ADAF Transitions} 

\author{Wei-Min Gu $^{1}$ and Ju-Fu Lu$^{1,2,3}$}
\altaffiltext{1}{Center for Astrophysics, University of Science and
Technology of China, Hefei, Anhui 230026, P. R. China;
gwm@mail.ustc.edu.cn}
\altaffiltext{2}{National Astronomical Observatories, Chinese Academy
of Sciences}
\altaffiltext{3}{Department of Physics, Xiamen University, Xiamen, Fujian
361005, P. R. China}

\begin{abstract}

We show that, unlike the results presented previously in the literature,
the transition from an outer Shakura-Sunyaev disk (SSD) to an
advection-dominated accretion flow (ADAF) is possible for large values
of the viscosity parameter $\alpha >0.5$. The transition is triggered
by thermal instability of a radiation-pressure-supported SSD. The
transition radius is close to the central black hole. We confirm our
qualitative prediction by actually constructing global bimodal SSD-ADAF
solutions. 
 
\end{abstract}


\newpage

\section {Introduction}

The model of bimodal accretion disk, which consists of a geometrically
thin, optically thick Shakura-Sunyaev disk (SSD) (Shakura \& Sunyaev 1973) 
as the outer part and a quasi-spherical, optically thin
advection-dominated accretion flow (ADAF) as the inner part, has been
quite successfully applied to black hole X-ray binaries and galactic
nuclei (see Narayan, Mahadevan, \& Quataert 1998 for a review). In this
model, the accretion flow switches from an SSD to an ADAF at a transition
radius $R_{tr}$. However, $R_{tr}$ was infered only by making plausible
assumptions on how it depends on the accretion rate $\dot M$. The precise
mechanism through which the SSD material is converted into an ADAF remains
a matter of debate, and no self-consistent global solution for such a
transition has been found so far. Indeed, Narayan, Kato, \& Honma (1997)
obtained a number of examples of global transonic ADAF solutions which
connect outward to geometrically thin disks. Lu, Gu, \& Yuan (1999)
recovered the whole class of such global ADAF-thin disk solutions. But
these authors did not take the local radiative cooling into account, thus
their solutions could not show the variation of the optical depth and
could not be regarded as bimodal SSD-ADAF solutions. On the other hand,
Chen, Abramowicz, \& Lasota (1997) considered the local radiative cooling
as provided by thermal bremsstrahlung, but did not find SSD-ADAF
solutions. Igumenshchev, Abramowicz, \& Novikov (1998) also adopted
bremsstrahlung cooling and found that only an outer
Shapiro-Lightman-Eardley (SLE) disk (Shapiro, Lightman \& Eardley 1976)
could smoothly match an inner ADAF. Such SLE-ADAF solutions are optically
thin everywhere. More impressive in this respect is the result of
Dullemond \& Turolla (1998, hereafter DT98). They concluded that the
SSD-ADAF transition was not permitted, and only an outer SLE disk could
match an inner ADAF or SSD. But, as they noticed, because of the thermal
instability of the SLE disk, such SLE-ADAF or SLE-SSD models most probably
do not exist in nature.

It would be a pity if such a promising bimodal SSD-ADAF model could not
be actually constructed. In this {\sl Letter} we also use an argument
based on energetic considerations which is similar to DT98, but find that
the SSD-ADAF transition is possible for large values of viscosity
parameter. We confirm our qualitative prediction by presenting a numerical
example of the global SSD-ADAF bimodal solution.
 
\section {Equations}

The system of equations adopted here, which is similar to DT98, makes the
following standard assumptions: \\ 
(1): The vertical half-thickness of the disk is expressed as $H=c_s/\Omega
_K$, where $c_s=(p/\rho )^{1/2}$ is the isothermal sound speed, with $p$
and $\rho $ being the total pressure and the mass density at the
equatorial plane, respectively, and $\Omega _K$ is the Keplerian angular
velocity calculated by using the Newtonian potential, $\Omega
_K=(GM/R^3)^{1/2}$. \\ 
(2): The kinematic viscosity coefficient is expressed as $\nu =\alpha 
c_sH$. \\ 
(3): $p$ is the sum of gas and radiation pressure, $p=p_g+p_r$.
$p_g=\rho {\cal R}(T_i+T_e)$, where $T_i$ and $T_e$ are the ion
temperature and the electron temperature, respectively, and 
$T_e=$min$(T_i, 6\times 10^9K)$. $p_r=Q_{rad}(\tau +2/\sqrt{3})/4c$, where
$Q_{rad}$ is the radiative cooling rate and $\tau =\kappa \rho H$ is the
total optical depth. \\ 
(4):The opacity $\kappa $ is the sum of electron scattering and
absorption opacity, $\kappa =\kappa _{es}+\kappa _{abs}$, where $\kappa
_{es}=0.34 cm^2g^{-1}$ and $\kappa _{abs}=0.27\times 10^{25}\rho 
T_e^{-3.5} cm^2g^{-1}$. 

Introducing the radial velocity $\upsilon _R$, the angular velocity
$\Omega $, and the surface density $\Sigma =2H\rho $, the continuity,
radial momentum, azimuthal momentum, and energy equations take the form
(e.g., Narayan \& Yi 1994; DT98)
\begin{eqnarray} 
\dot M=-2\pi \Sigma R\upsilon _R \ ,
\end{eqnarray} 
\begin{eqnarray}
\upsilon _R\frac{d\upsilon _R}{dR}+(\Omega _K^2-\Omega ^2)R+\frac 1\rho 
\frac{dp}{dR}=0 \ ,
\end{eqnarray}
\begin{eqnarray}
\Sigma R\upsilon _R\frac{d(\Omega R^2)}{dR}=\frac d{dR}(\Sigma \nu R^3
\frac{d\Omega }{dR}) \ ,
\end{eqnarray}
\begin{eqnarray}
Q^{+}=Q^{-}=Q_{adv}+Q_{rad} \ .
\end{eqnarray}
The viscous heating $Q^{+}$ has the usual expression
\begin{eqnarray}
Q^{+}=\nu \Sigma (R\frac{d\Omega }{dR})^2 \ .
\end{eqnarray}
The advective cooling is expressed as
\begin{eqnarray}
Q_{adv}=\Sigma \upsilon _RT\frac{ds}{dR}=\Sigma
\upsilon _R(\frac{1}{\gamma -1}\frac{dc_s^2}{dR}-\frac {c_s^2}{\rho
}\frac{d\rho}{dR}) \ .
\end{eqnarray}
The radiative cooling rate $Q_{rad}$ is calculated using a bridging
formula (DT98), 
\begin{eqnarray}
Q_{rad}=8\sigma T_e^4(\frac{3\tau }2+\sqrt{3}+\frac{8\sigma
T_e^4}{Q_{brem}})^{-1} \ .
\end{eqnarray}
Eq.(7) is valid in both optically thin and optically thick regimes. 
The bremsstrahlung cooling is given by (Abramowicz et al. 1995)
\begin{eqnarray}
Q_{brem}=1.24\times 10^{21}H\rho ^2T_e^{1/2}ergs \
s^{-1}cm^{-2} \ .
\end{eqnarray}

Abramowicz et al. (1995) and Chen et al. (1995, hereafter Chen95) obtained
a unified $\Sigma -\dot M$ picture of accretion flows around black holes
with an additional assumption that the disk always rotates at $\Omega
=\Omega _K$. However, it is known that slim disks (Abramowicz et al.
1988) and ADAFs are quite sub-Keplerianly rotating. We therefore assume:
$\Omega =\omega \Omega _K (0\leq \omega \leq 1)$, where the parameter
$\omega $ is obtained from the self-similar solution (Narayan \& Yi,
1994). We fix $\gamma =1.5$ and adopt the self-similar method, then the
equations (2), (3), (5) and (6) are reduced to the following algebraic
form:  
\begin{eqnarray}
\frac {1}{2}\upsilon _R^2+\frac {5}{2}c_s^2+(\omega ^2-1)\Omega _K^2R^2=0
\ ,  
\end{eqnarray}
\begin{eqnarray}
\upsilon _R=-\frac {3}{2}\frac {\nu}{R}=-\frac {3}{2}\alpha c_s\frac
{H}{R} \ ,
\end{eqnarray}
\begin{eqnarray}
Q^{+}=\frac {3}{4\pi }\dot M\Omega ^2 \ ,
\end{eqnarray}
\begin{eqnarray}
Q_{adv}=\frac {1}{4\pi }\frac{\dot Mc_s^2}{R^2} \ .
\end{eqnarray}

From Eq.(9) one easily obtains: $h=H/R=c_s/\Omega _KR<\sqrt{2/5}\approx 
0.63$. We use a parameter $f=Q_{adv}/Q^{+}=h^2/3\omega ^2$ to measure the 
degree to which the flow is advective. If the radial velocity is
neglected, Eq.(9) is simplified to: $(\omega ^2-1)+\frac {5}{2}h^2=0$.
For radiative cooling-dominated flows ($f\approx 0$) such as SSDs or SLE
disks, $\omega \approx 1$, and $h\approx 0$; while for advection-dominated
flows ($f\approx 1$) such as slim disks or ADAFs, $\omega \approx \sqrt
{2/17}$, and $h\approx \sqrt {6/17}$. Thus the assumption $\Omega =\omega
\Omega _K$ is more reasonable as it is valid for both radiative
cooling-dominated and advection-dominated accretion flows. 

\section {SSD-ADAF Solutions}

Abramowicz et al. (1995) first obtained a unified $\Sigma -\dot M$
picture for accretion flows at a fixed radius $R$ in the case of low
viscosity, which includes four classes of solutions, namely SSDs, SLE
disks, slim disks and ADAFs. Chen95 found that two types of $\Sigma 
-\dot M$ picture should exist, which are separated by a critical viscosity
parameter $\alpha _{crit}$. We recover here these two types of $\Sigma
-\dot M$ picture with our assumption $\Omega =\omega \Omega _K$, which
are given in Fig.1, (a) for $\alpha <\alpha _{crit}$ and (b) for
$\alpha >\alpha _{crit}$. 
The solid lines in the figures represent thermal equilibrium solutions, 
i.e. with $Q^{+}=Q^{-}$. In Fig.1(a), the right $S$-shaped curve consists
of three branches, of which the lower one is for gas-pressure-supported
SSDs, the middle one for radiation-pressure-supported SSDs, and the upper
one for slim disks; while the left curve consists of two branches, of 
which the lower one is for SLE disks, and the upper one for ADAFs. In
Fig.1(b), the straight line is for ADAFs and slim disks, while the 
$n$-shaped curve consists of three branches, of which the two branches
on the right are the same as the middle and lower branches of the
$S$-shaped curve in Fig.1(a), and the branch on the left is for SLE disks.
The unstable branches are those which have $Q^+>Q^-$ above and $Q^+<Q^-$
below, while the stable branches are just opposite. Thus
gas-pressure-supported SSDs, slim disks and ADAFs are thermally stable,
whereas radiation-pressure-supported SSDs and SLE disks are thermally
unstable.

DT98 investigated the thermal instability in Fig.1(a), and argued that
the SSD-ADAF transition is not permitted. We agree with DT98 that such 
a transition is indeed prohibited in Fig.1(a). The arrows in Fig.1(a) show
a limit-cycle behavior resulting from the thermal instability of a
radiation-pressure-supported SSD. 

However, the other type of picture as shown in Fig.1(b) was ignored by
DT98. Because a bridging formula like Eq.(7) is used, the optically thick,
high $\dot M$ slim disk solution and the optically thin, low $\dot M$ ADAF
solution can be described by a single line in Fig.1(b), the line extends
over the entire range of $\dot M$ without break. It is this feature that
makes the SSD-ADAF transition possible. A thermal disturbance on a
radiation-pressure-supported SSD can trigger the flow to behavior in the
following way: The flow first jumps to a slim disk solution and becomes
thermally stable. But because the accretion rate $\dot M$ does not match
that of the outer SSD, then the slim disk must evolve into an ADAF, for
which $\dot M$ matches that of the outer SSD. The whole process is
indicated by the two arrows in Fig.1(b). An SSD-ADAF transition is
realized.

To confirm this qualitative prediction, we now go on to present our
numerical models for bimodal SSD-ADAF disks. The general thermal
instability condition is
\begin{eqnarray}
\left( \frac{\partial Q}{\partial T}\right) _\Sigma =\left( \frac{\partial
Q^{+}}{\partial T}\right) _\Sigma -\left( \frac{\partial Q^{-}}{\partial
T}\right) _\Sigma >0
\end{eqnarray}
We denote $\beta =p_g/(p_g+p_r)$, and $\lambda =\kappa _{abs}/(\kappa
_{es}+\kappa _{abs})$. In privious researches on SSDs the disk was usually
divided into three separate regions, for which (from the inner to the
outer) the two parameters $(\beta ,\lambda )$ are (0,0), (1,0) and (1,1),
respectively. Here we let both $\beta $ and $\lambda $ vary continuously
from 0 to 1, thus the SSD solution obtained will smoothly extend over
these three regions. In our formulation Eq.(13) takes an explicit form
\begin{eqnarray}
\delta =4-10\beta -7.5\lambda -0.5\beta \lambda>0 \ .
\end{eqnarray}
When the opacity is dominated by the electron scattering, i.e. $\lambda 
=0$, this condition is reduced to the usual expression: $\beta <0.4$.

Once the viscosity parameter $\alpha$ and the relative accretion rate 
$\dot m$ ($\dot m=\dot M/M_{Edd}$ with $M_{Edd}$ being the Eddington
accretion rate) are given, we can search for the global SSD-ADAF solution
by the following steps:\\ 
(i) Provided that the given viscosity parameter $\alpha >0.5$, a critical
radius $R_c$ exists. The $\Sigma -\dot M$ picture is of the type of
Fig.1(a) for $R>R_c$, and of the type of Fig.1(b) for $R<R_c$. We first
calculate the value of $R_c$ which corresponds to the given $\alpha $.\\
(ii) We solve the equations of SSD inward from an outer boundary 
$R_{out}=2000R_g$. The SSD solution breaks off when the thermal
instability condition Eq.(14) is met at a certain radius $R_b$. We
calculate the value of $R_b$ which corresponds to the given $\alpha $
and $\dot m$. If the condition Eq.(14) is never met for the given $\alpha 
$ and $\dot m$, then the SSD solution is thermally stable everywhere. \\
(iii) If $R_b<R_c$, the SSD-ADAF transition will occur at $R_{tr}=R_b$. 
We can obtain the self-similar ADAF solution inside $R_{tr}$. Combining 
with the SSD solution outside $R_{tr}$, we obtain the whole global
SSD-ADAF solution. On the contrary, if $R_b>R_c$, the limit-cycle behavior
will occur at $R_b$, and the SSD-ADAF transition will not occur.  

In Fig.2 we show how the behavior of an original SSD flow depends on 
$\alpha $ and $\dot m$. It is seen that three possible cases, namely the
stable SSD, the limit cycle behavior, and the SSD-ADAF transition, each
occupy a certain region of the $\alpha -\dot m$ parameter space, and the
SSD-ADAF transition occurs for $\alpha >0.5$ and a definite range of $\dot
m$. To construct numerically a global bimodal SSD-ADAF solution, we
choose $\alpha =0.8$. We then obtain the corresponding critical radius
$R_c=12.0R_g$, and find from Fig.2 that the SSD-ADAF transition occurs for
$0.03<\dot m<0.17$. The SSD is stable everywhere for $\dot m<0.03$, and
the limit-cycle behavior appears for $\dot m>0.17$. An example of the
global SSD-ADAF solution for $\dot m=0.1$ is presented in Fig.3. The SSD
becomes unstable at $R_b=8.0R_g$. The figure gives a $log(\Sigma)-log(\dot
m)$ picture at $R=8.0R_g$, which belongs to the type of Fig.1(b). The
arrows indicate the transition from the SSD solution (filled circle) to an
ADAF solution (filled square).     

\section {Discussion}

In this {\sl Letter} we show that the thermal instability of a
radiation-pressure-supported SSD can possibly trigger two different kinds
of behavior of the flow, namely the limit-cycle and the SSD-ADAF
transition. For low values of viscosity parameter $\alpha <0.5$, only the
limit-cycle behavior can occur; while for large values $\alpha >0.5$,
either of the two kinds of behavior can, and which one is actually
realized is determined by $\dot m$. We use two parameter, $\beta $
and $\lambda $, to smoothly connect the three usually separated regions of
an SSD, so that the exact position of the SSD-ADAF transition can be
found, and the global bimodal SSD-ADAF solution can be obtained.  

The range of $\alpha $ values necessary for the SSD-ADAF transition to
occur should depend on, among other things, radiative cooling mechanisms
adopted. Chen95 used two bridging formulae of $Q_{rad}$ to calculate 
$\alpha _{crit}$ for a fixed radius $R$, one of which is introduced by
Narayan \& Yi (1995), who considered bremsstrahlung and synchrotron
cooling, and Comptonization; and the other of which is introduced by
Wandel \& Liang (1991), who considered bremsstrahlung cooling with
Comptonization only. They found that the value of $\alpha _{crit}$
calculated with the former formula (i.e. with more sources of radiative
cooling) is higher than that calculated with the latter one. For example,
for $R=5R_g$, they had $\alpha _{crit}=0.41$ with the former formula, 
and $\alpha _{crit}=0.245$ with the latter one. Since in bridging formula
Eq.(7) only bremsstrahlung cooling is taken into account, we guess that
if other sources of radiative cooling such as synchrotron emission and
Comptonization are included, the lowest required value of $\alpha $ for 
the SSD-ADAF transition would become larger than 0.5, i.e. the necessary
condition on $\alpha $ for the existence of bimodal SSD-ADAF disk would 
seem even more restrictive. It is unclear yet if there are other factors
which could help to relax this condition on $\alpha $. However, the good
thing for the SSD-ADAF transition is that the larger the allowed value
of $\alpha $ is, the wider the corresponding range of $\dot m$ is, as seen
clearly from Fig.2.  

The SSD-ADAF transition found in this {\sl Letter} results from the
instability of the radiation-pressure-supported region, i.e. the inner
part of SSD. Thus the transition position is close to the central black
hole, as we have calculated here. On the other hand, observations of
some black hole X-ray binaries (Narayan et al. 1998 and references
therein) seem to imply a SSD-ADAF transition radius $R_{tr}\sim 10^4R_g$,
which is definitely in the very outer region of SSD. However, it is well
known that the outer region of SSD is gas-pressure-supported and is both
thermally and viscously stable. In our opinion, it is difficult to see 
from the theoretical point of view how to convert a gas-pressure-supported 
SSD directly into an ADAF. The apparent conflict between present
observations and theories remains an unsolved issue.

\newpage

\figcaption[f1a.ps,f1b.ps]{Two types of $log(\Sigma )-log(\dot M)$ 
picture for accretion flows at a fixed radius. The Solid lines mark the
different branches of solutions. The arrows show the behavior of the 
flow resulting from the thermal instability of a
radiation-pressure-supported SSD.
(a) $\alpha <\alpha_{crit}$, a limit-cycle occurs.
(b) $\alpha >\alpha_{crit}$, an SSD-ADAF transition occurs.
\label{Fig-1}}

\figcaption[f2.ps]{Dependence of the possible behavior of an original
SSD flow on $\alpha $ and $\dot m$.
\label{Fig-2}}

\figcaption[f3.ps]{An example of global SSD-ADAF solution with $\alpha 
=0.8$, and $\dot m=0.1$. The $log(\Sigma )-log(\dot m)$ picture is for
the accretion flow at $R=8.0R_g$, where the SSD becomes thermally
unstable. The arrows show the transition from the SSD (filled circle) to
an ADAF (filled square). The dotted line corresponds to the limit:
$H/R=0.63$. 
\label{Fig-3}}

\newpage

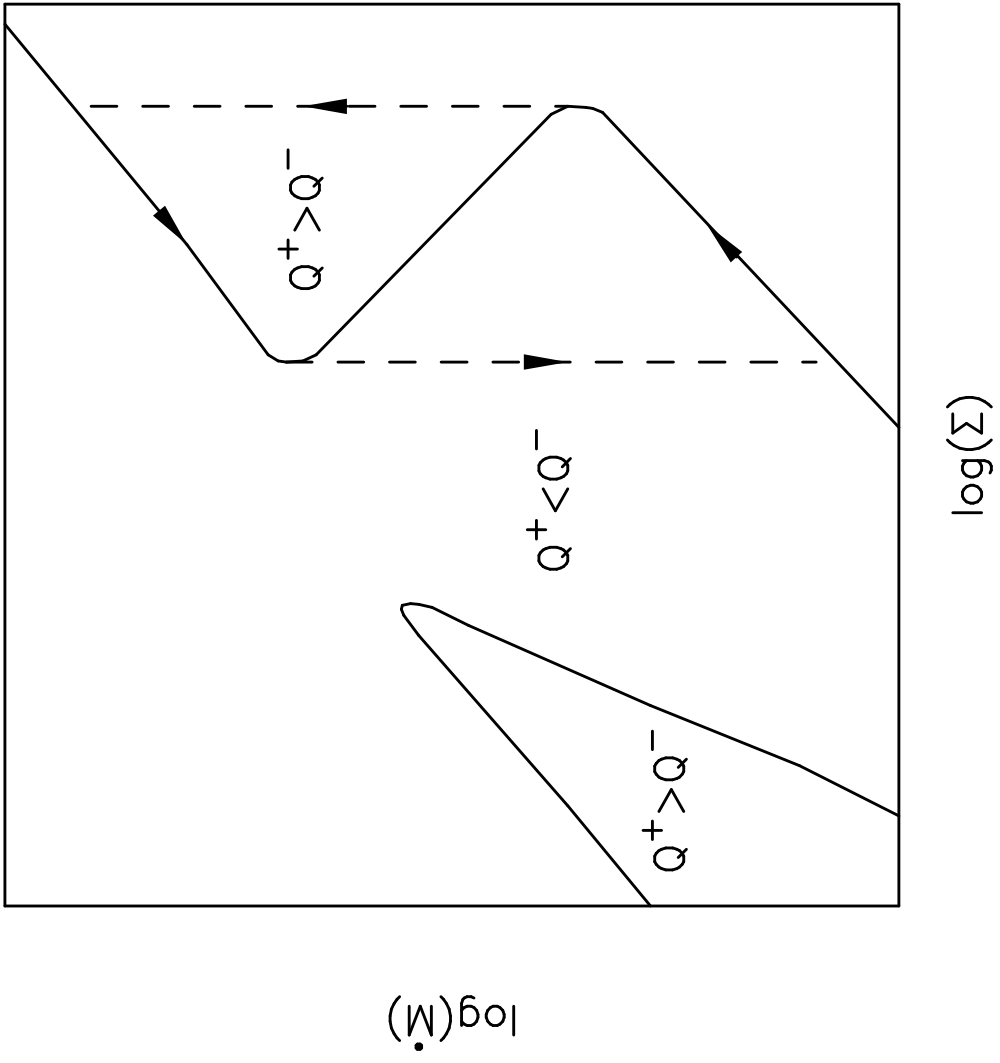
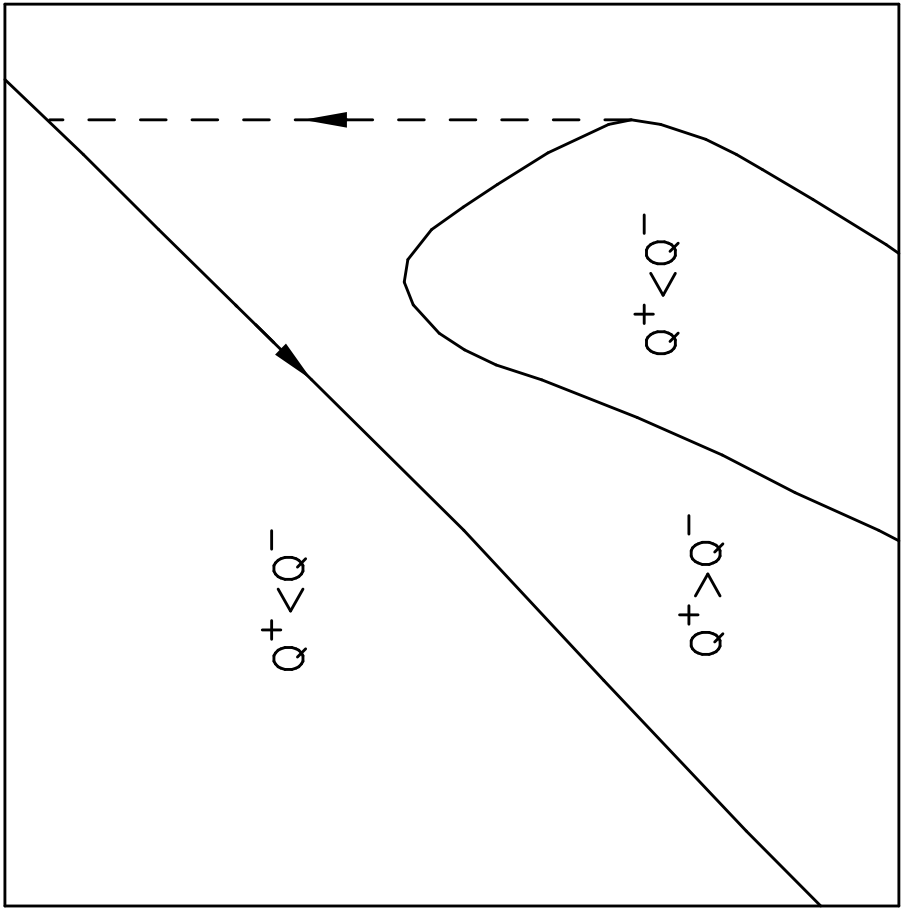
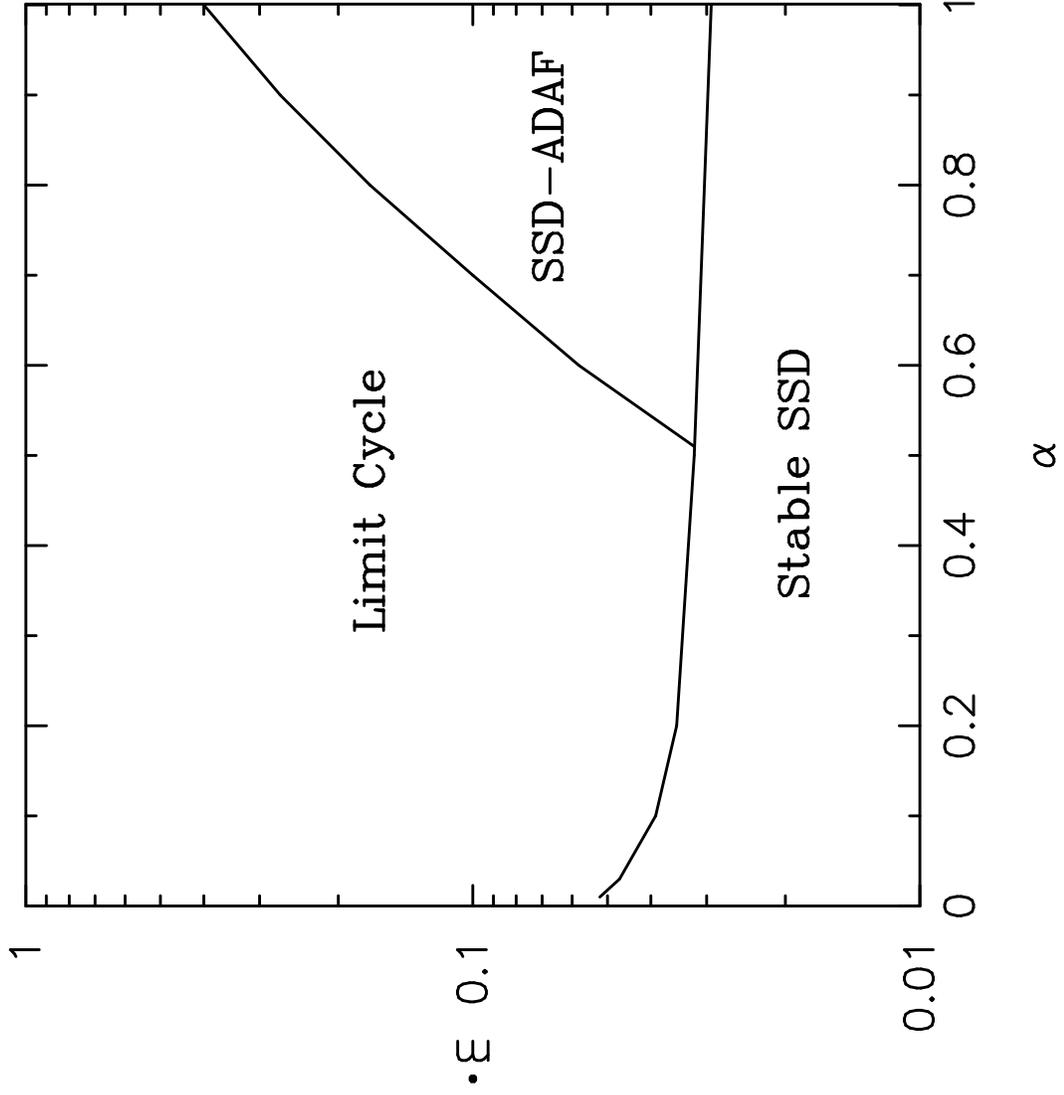
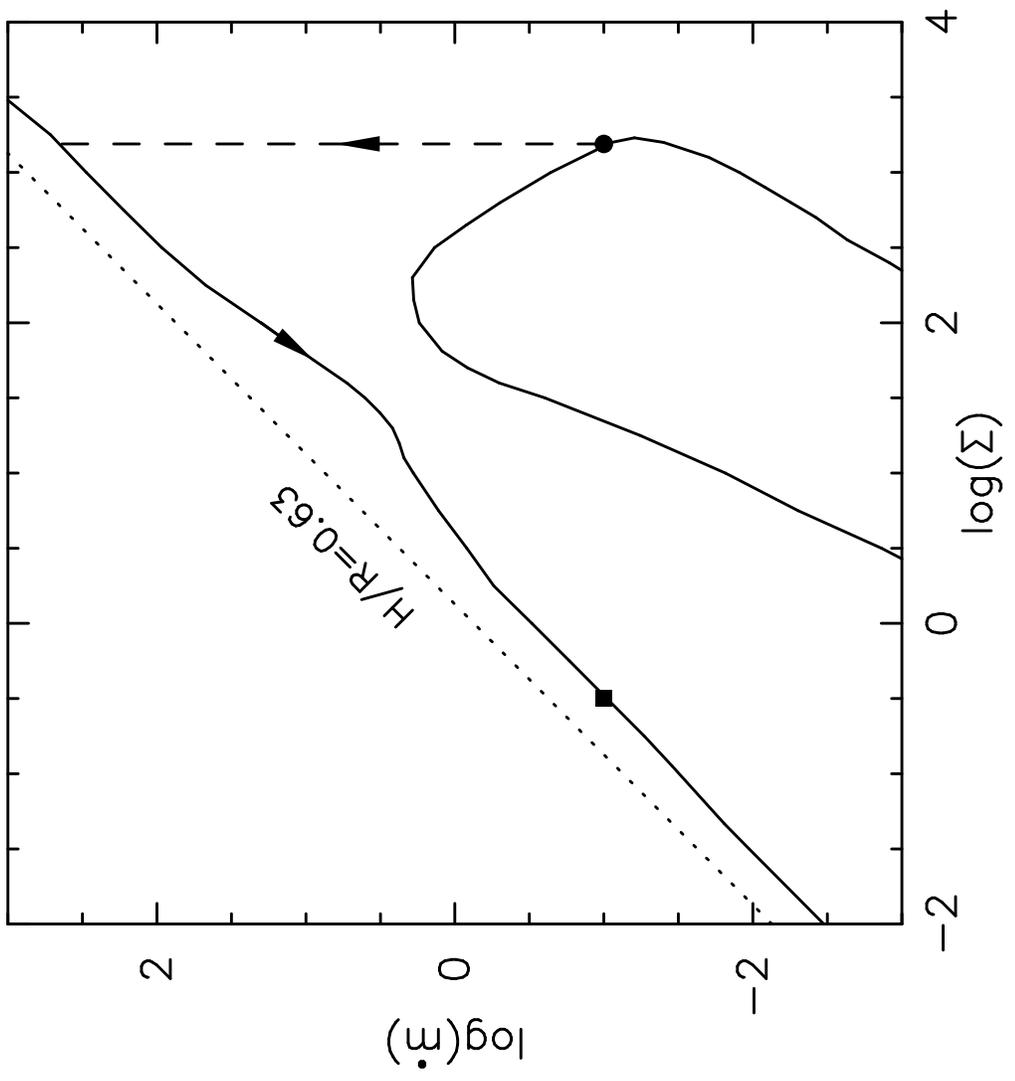
\end{document}